\documentclass[fleqn,10pt]{wlscirep}
\usepackage[utf8]{inputenc}
\usepackage[T1]{fontenc}
\usepackage{tabularray}
\usepackage{fontawesome5}
\usepackage{xspace}
\usepackage{lineno}

\definecolor{commentred}{RGB}{240,0,0} 
\definecolor{commentgreen}{RGB}{0,200,0} 

\def\sd#1{\color{black} {\scriptsize $\pm$ \hspace{-1.4mm} #1} \color{black}}

\def\inc#1{\textcolor{commentgreen} {\scriptsize $\uparrow$ \hspace{-1.2mm}  #1} \color{black}}
\def\dec#1{\textcolor{commentred} {\scriptsize $\downarrow$ \hspace{-1.2mm}  #1} \color{black}}

\newcommand{\mname}{UniCoN\xspace}

\title{\mname: Universal Conditional Networks for Multi-Age Embryonic Cartilage Segmentation with Sparsely Annotated Data}

\author[1,+,*]{Nishchal Sapkota}
\author[1,+]{Yejia Zhang}
\author[1]{Zihao Zhao}
\author[1]{Maria Gomez}
\author[2]{Yuhan Hsi}
\author[2]{Jordan A. Wilson}
\author[2]{Kazuhiko Kawasaki}
\author[3]{Greg Holmes}
\author[4]{Meng Wu}
\author[3,4]{Ethylin Wang Jabs}
\author[2]{Joan T. Richtsmeier}
\author[2]{Susan M. Motch Perrine}
\author[1]{Danny Z. Chen}

\affil[1]{Department of Computer Science and Engineering, University of Notre Dame, Notre Dame, IN 46556, USA}
\affil[2]{Department of Anthropology, The Pennsylvania State University, University Park, PA 16802, USA}
\affil[3]{Department of Genetics and Genomic Sciences, Icahn School of Medicine at Mount Sinai, Icahn Medical Institute, New York, NY 10029, USA}
\affil[4]{Department of Clinical Genomics, Mayo Clinic, Rochester, MN 55905, USA}
\affil[*]{\hspace{0.2mm}\faEnvelope:\hspace{0.5mm}nsapkota@nd.edu}
\affil[+]{Indicates equal contributions.}

\begin{abstract}
Osteochondrodysplasia, affecting 2-3\% of newborns globally, is a group of bone and cartilage disorders that often result in head malformations, contributing to childhood morbidity and reduced quality of life.
Current research on this disease using mouse models faces challenges since it involves accurately segmenting (precisely delineating) the developing cartilage in 3D micro-CT images of embryonic mice. Tackling
this segmentation task with deep learning (DL) methods is laborious due to the big burden of manual image annotation, expensive due to the high acquisition costs of 3D micro-CT images, and difficult due to embryonic cartilage's complex and rapidly changing shapes.
While DL approaches have been proposed to automate cartilage segmentation, most such models have limited accuracy and generalizability, especially across data from different embryonic age groups.
To address these limitations, we propose novel DL methods that can be adopted by any DL architectures -- including Convolutional Neural Networks (CNNs), Transformers, or hybrid models -- which effectively leverage age and spatial information to enhance model performance. 
Specifically, we propose two new mechanisms, one conditioned on discrete age categories and the other on continuous image crop locations, to enable an accurate representation of cartilage shape changes across ages and local shape details throughout the cranial region.
Extensive experiments on multi-age cartilage segmentation datasets show significant and consistent performance improvements when integrating our conditional modules into popular DL segmentation architectures. On average, we achieve a 1.7\% Dice score increase with minimal computational overhead and a 7.5\% improvement on unseen data. These results highlight the potential of our approach for developing robust, universal models capable of handling diverse datasets with limited annotated data, a key challenge in DL-based medical image analysis. \\


\end{abstract}
\begin{document}
\flushbottom
\maketitle

\thispagestyle{empty}

\noindent 

\vspace{-10mm}
\section*{Introduction}  
\label{sec:1}

Osteochondrodysplasia is a clinically heterogeneous group of more than 100 genetic disorders causing developmental defects in connective tissues, including cartilage and bones. The World Health Organization estimated that 2-3\% of human newborns are affected by such birth defects~\cite{who}.
A particularly concerning aspect of osteochondrodysplasia is its association with cranial dysmorphology where malformations of the chondrocranium may lead to morphological defects in the bony skull during embryogenesis. 
The chondrocranium is the cartilaginous structure that encases and protects the developing brain and sense organs, providing initial support during embryogenesis~\cite{pitirri2022meckel, perrine2023embryonic}. 
Development of the chondrocranium and Meckel’s cartilage of the lower jaw, which precedes the formation of the dentary bone, plays a crucial role in normal craniofacial development. 
Given that embryonic cartilage growth is a highly complex and tightly regulated process, any disruption can lead to severe phenotypic abnormalities. Recent studies~\cite{perrine2023embryonic, sapkota2024conunetr,hao2020cartilage} tackled the challenge of accurate delineation and segmentation of cartilaginous structures of the head to better understand the underlying mechanisms of abnormal development and potentially reveal new therapeutic targets.
Due to the invasive nature of these studies, cross-sectional techniques using contrast-enhanced micro-computed tomography (micro-CT) imaging of embryonic mouse models of human conditions are favored for visualization of embryonic cartilage development.
However, extracting precise cartilage regions in these large three-dimensional (3D) images is difficult due to the fast-changing cartilage morphology during early development and the prohibitive costs of expert annotation.

\begin{figure*}[h!] 
\begin{center} 
\includegraphics
[width=0.9\linewidth]
{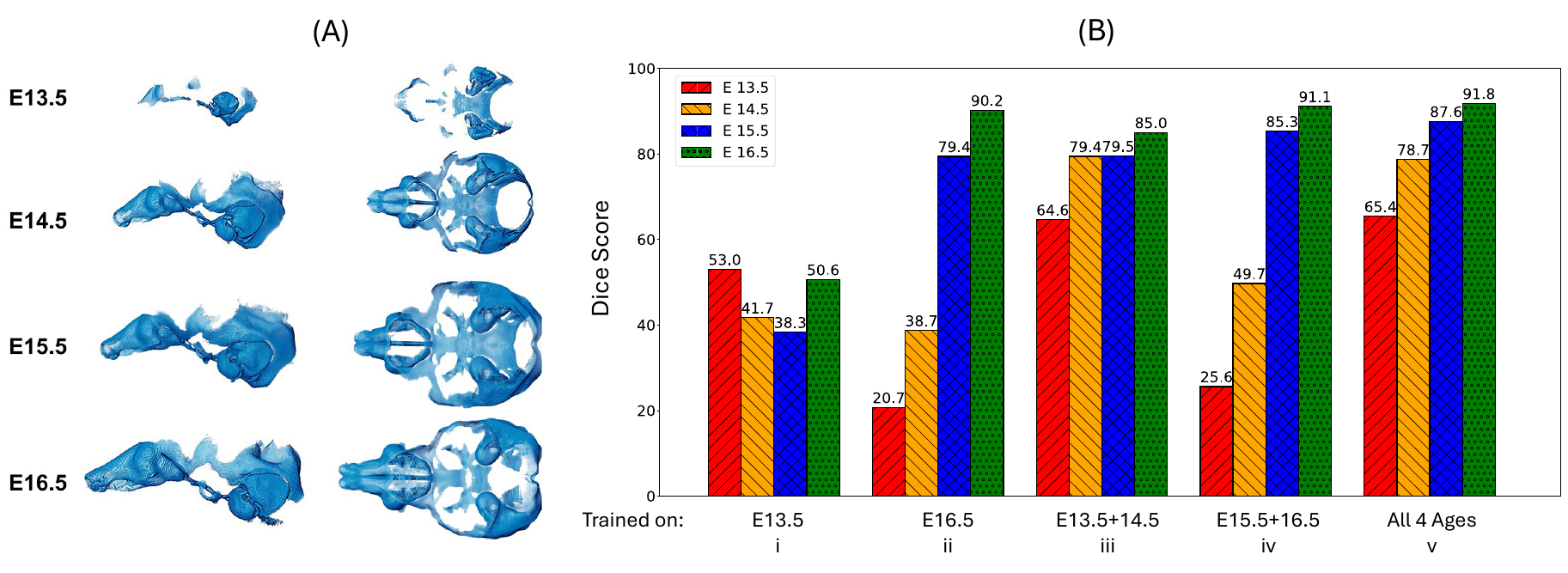} 
\caption{\label{MAV}
(A) Visualizing the morphological variations in mouse cartilage across four embryonic ages measured as days after fertilization. (B) The performance of Res2Unet*, a popular deep neural network model for segmentation, on a multi-age cartilage dataset (C57), using different ages for training. \vspace{-5mm}
}
\end{center}
\end{figure*}

Recently, many works have adopted deep neural networks to automate bone and cartilage segmentation with data-driven approaches.
Most such works focus on bone and cartilage segmentation in the knee \cite{shapePriorsCart, li2023sdmt, poistionPriorClusterKneeCartSeg, lin2022calibrating, liu2022isegformer, prasoon2013deep} or hip \cite{zeng2020entropyHip} region from magnetic resonance images.
However, the study of embryonic cartilage segmentation in 3D micro-CT images remains limited. 
Three main challenges hamper large data-driven studies on this task. 
First, micro-CT per specimen is expensive, time-consuming to capture, and thus limited in number.
Second, labeling the cartilage in 3D images is costly, laborious, and requires expert anatomists. 
Third, cartilage presents sizable morphological and geometric diversity during embryonic development, making them difficult to model. 
An example of morphological variations in mouse cartilage structures across embryonic (E) ages is illustrated in Figure \ref{MAV}-(A) \cite{perrine2022elife}, where one can see discontinuous and porous cartilage structures, ambiguous cartilage boundaries, and large variations in structure thickness. 
Previous works on embryonic-stage cartilage segmentation often did not use multi-age data \cite{hao2020cartilage, matula2022resolving}, while the ones that utilized multi-age data (e.g., Blumer et al.~\cite{blumer2004cartilage}) did not adopt data-driven approaches and relied on staining to delineate cartilage structures. 
Another common strategy to address structure variations involves training a separate model for each age group~\cite{perrine2023embryonic, perrine2022elife, pitirri2022meckel}.
However, this approach requires larger datasets, intensive annotation efforts by experts, and higher computation costs. 
Also, models trained on one age group may overfit the data segment on which it was trained; Figure \ref{MAV}-(B)-ii illustrates this scenario where a DL-based segmentation model, Res2Unet*, was trained on data from age group E16.5 (meaning 16.5 embryonic days after fertilization) and performs poorly when used for prediction on earlier ages such as E13.5 and E14.5.

Alternatively, a single model trained on a combined multi-age dataset, or joint training, would result in a reduction of annotation and training efforts, along with increased performance due to the advantage of an expanded training set that leverages structural similarities across ages.
Figure \ref{MAV}-(B)-v illustrates this observation where a Res2Unet* is trained jointly on all 4 ages, is then applied to multi-age test sets, and achieves better performances. 
However, naively combining multiple ages into a single dataset does not adequately exploit the large structural variances across ages and the subtle shape variations across scans.
Jointly trained models are often biased toward the structural characteristics of one age group over another. For instance, they often under-segment the structures for later ages or over-segment the structures at earlier ages~\cite{sapkota2024conunetr} (see Figure \ref{MAV}-(B)-iv which shows poor performance on earlier age groups when trained on older ages).

To better capture both local cartilage details and global variations across ages, we aim to leverage age and spatial information that is freely available in our multi-age 3D dataset. 
Our main idea is to inject this information at various pertinent points in the segmentation model to enrich image features and ultimately improve cartilage segmentation. 
To achieve this, we introduce feature enrichment modules that are conditioned on discrete ages (e.g., E13.5, E14.5, E15.5, or E16.5) and on continuous spatial information (e.g., encoding the position of each crop with respect to the whole 3D image) to distill helpful contextual information beyond what is present in the input images. 
This concept of incorporating non-visual information has previously been explored in medical image analysis.
For instance, several works~\cite{yan2023sparseWSI, zhang2023ihcsurvWSI, huang2021integrationWSI} leveraged the Transformer architecture~\cite{transformer, vit} by feeding non-visual information as a learnable token to represent discrete categories or converting the information to spatial embeddings to be combined with input tokens. 
Similar to our approach, ConUNETR~\cite{sapkota2024conunetr} previously tackled embryonic cartilage segmentation using ``Age" tokens to distill information on structural similarities and disentangle dissimilarities across embryonic ages. 
This model exhibited state-of-the-art performance and improved generalizability than naively joint training schemes. 
However, ConUNETR still poses disadvantages by adopting a data-hungry encoder in a task that often involves sparse annotated data availability (as shown in Table~\ref{main_table}) by using the Transformer architecture. 

In this paper, we propose a novel architecture-agnostic strategy that any DL segmentation models can adopt to jointly train on multi-age cartilage datasets. 
Our strategy is based on new modules inserted at multiple points in the segmentation model, which enrich features by distilling information from discrete age categories and continuous spatial information.
The segmentation models built using these modules --- \mname (for Universal Conditional Networks) --- work very well in sparse annotation settings and incur minimal computational overhead. 
We demonstrate the effectiveness of our approach on multi-age cartilage segmentation datasets, resulting in notable and consistent performance gains when integrating our conditional modules into six popular DL segmentation architectures. 
On average, we achieve 1.7\% Dice score improvements compared to the original architecture performances and a 7.5\% Dice score improvement when applying our approach to unseen data.

Our main contributions are as follows: 
(1) We introduce a novel universal conditioning strategy based on age segments that are compatible with any DL segmentation architectures (e.g., CNNs, Transformers, hybrid); 
(2) we judiciously design the new conditional modules to augment model features using rich prior information with small computational overhead;
(3) we achieve state-of-the-art performance in extensive experiments on a challenging cartilage segmentation dataset with four age groups, significantly outperforming ConUNETR~\cite{sapkota2024conunetr} despite using significantly sparser annotated data;
(4) we demonstrate superior zero-shot transfer performance on unseen data.

\section*{Methods}  
\label{sec:2}

\begin{figure*}[h!] 
\begin{center} 
\includegraphics[width=0.7\linewidth]{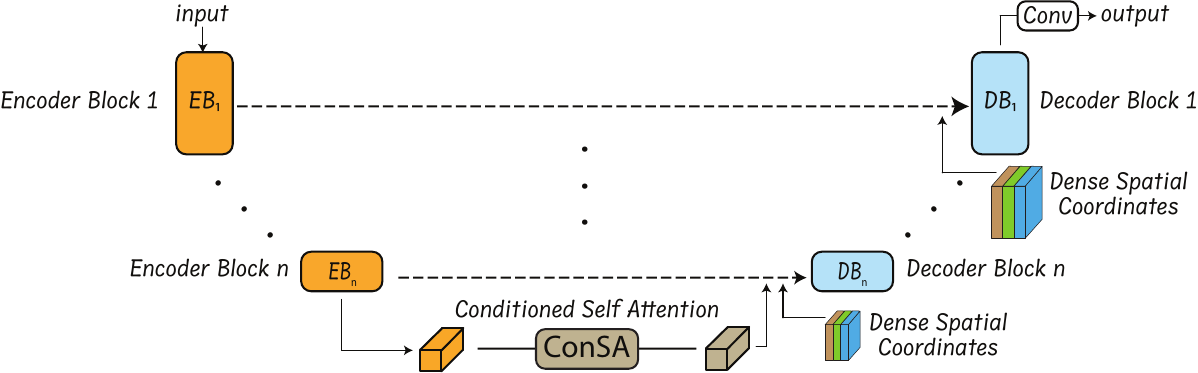} 
\caption{\label{main}
An overview of our proposed \mname. Conditional modules added to U-Net-shaped segmentation models:
Conditional self-attention (ConSA) is placed at the bottleneck, and hierarchical dense spatial coordinates (HDSCs) are added to the decoder blocks. The flexibility of this design allows our approach to be architecture-agnostic. \vspace{-5mm}
}
\end{center}
\end{figure*}

Our main idea is to leverage joint training, or training on combined data from multiple embryonic ages simultaneously, to capture shared cartilage structure characteristics while introducing new conditional components to improve the model representation of structure differences.
These conditional components incorporate discrete age and continuous spatial information to enhance the representational power of any base segmentation model. We adopt U-Net~\cite{unet}-shaped models -- the most popular and performant model structure for segmentation -- which consist of encoders (CNN, Transformer, or Hybrid), decoders (CNN), and skip connections between the encoder and decoder. However, since we add our conditional components at the bottleneck and various decoder stages (see Figure \ref{main}), our design is architecture-agnostic and can be seamlessly integrated with CNN-based, Transformer-based, or other hybrid segmentation backbones. We call these segmentation models \mname (Universal Conditional Networks).


Our conditional components consist of two modules:
(1) a conditional self-attention (ConSA) module that contains a self-attention mechanism that conditions on age and crop-level spatial information, and 
(2) a hierarchical dense spatial coordinates (HDSC) concatenation that is attached to the decoder blocks at multiple scales and concatenates dense spatial information in a channel-wise manner to multi-scale features obtained from the encoder.
Empirically, we found that adding a single ConSA to the encoder's lower resolution outputs and adding HDSCs to each decoder block's input immediately after the encoder skip connections performed the best. 
The following subsections will describe these components. 

\subsection*{Conditioned Self-Attention}

Inspired by ConUNETR~\cite{sapkota2024conunetr}, we introduce the conditioned self-attention (ConSA) module to better distill age-specific cartilage structure information during the feature encoding process.
Unlike ConUNETR which introduced age tokens at the beginning of the encoder layers, our ConSA uses age information and additionally spatial features corresponding to image features, at the lowest resolution output of the encoder stage.
This design improves model performance in sparse annotated data settings as we are decoupling the encoder's feature learning process from the incorporation of critical non-visual information. 
In the segmentation pipeline, incorporating critical non-visual information (age and spatial information) after learning image features enhances the segmentation model's contextual awareness without adding model complexity to the feature learning process which risks overfitting. 
As a result, ConSA can output age-relevant and spatially-aware image features conditioned on both the respective age group of the samples and their spatial context, improving multi-age segmentation performance.


\begin{figure*}[h!] 
\begin{center} 
\includegraphics
[width=0.95\linewidth]
{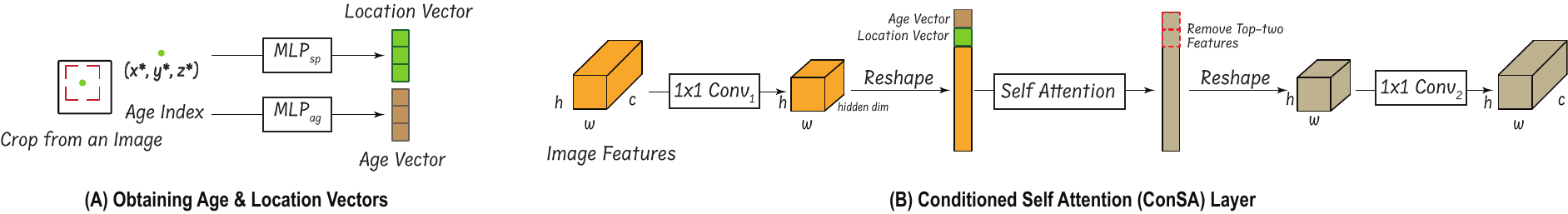} 
\caption{\label{conSA}
(A) Obtaining age and location vectors from an input image. (B) Obtaining image features conditioned on age and spatial information by applying the conditioned self-attention (ConSA) layer to the image features outputted at the end of the encoder stage. \vspace{-7mm}
}
\end{center}
\end{figure*} 
Figure \ref{conSA}-(A) illustrates the conditioning process and how age and spatial information features are obtained and encoded from an input image. 
Given a $z^{th}$ image slice $s_z$ of dimension $H \times W$ from a volume (with totally $Z$ slices) belonging to the age group $A_i$, where $A_i \in \{0,1,2,3\}$ such that $\{\text{0: E13.5, 1: E14.5, 2: E15.5, 3: E16.5}\}$, we crop an $h \times w$ region in $s_z$ with randomly selected top ($t$), bottom ($b$), left ($l$), and right ($r$) locations. Next, we obtain the relative central coordinates $(x^*, y^*, z^*)$ of $s_z$, as follows:
\begin{align}  
\; \; \; \; \; \; \; \; \; \; \; \; \; \; \; \; \; \;
x^* = \frac{l+r}{2 \times W}, \; \; 
y^* = \frac{t+b}{2 \times H}, \; \;
z^* = \frac{z}{Z}.
\end{align}
The age index of the input image and the relative central coordinates $(x^*, y^*, z^*)$ are mapped using two separate multilayer perceptions (MLPs) to obtain the $age \; vector$ and $location \; vector$, both of dimension $hid\_dim$, as:
\begin{align}
\; \; \; \; \; \; \; \; \; \; \; \; \; \; \; \; \; \; Age \; Vector &= MLP_{ag} (A_i), \; \; \;
Location \; Vector = MLP_{sp} (x^*, y^*, z^*).
\end{align} 


The image crop is fed into the encoder of the segmentation model, and the lowest-resolution image features obtained at the end of the encoding stage are passed through the ConSA layer as shown in Figure \ref{conSA}-(B). For a given $c  \times h  \times w $ (channel $\times$ height $\times$ width) shaped features outputted by the encoder, we apply a $1 \times 1$ convolution block to obtain a $hid\_dim \times h \times w$ shaped features and flatten them to $(h \cdot w) \times hid\_dim$, where $hid\_dim$ corresponds to the dimension of the embedding that is used for each token. 
We concatenate the $age \; vector$ and $location \; vector$ at the top of the flattened image features. Next, a multi-headed self-attention processes the features, age and location vectors. 
The output features now contain embedded age and location information within each patch token. 
The patch tokens -- not including the $Age \; Vector$ and $Location \; Vector$ -- are then reshaped to $hid\_dim \times h \times w$ and passed through a $1 \times 1$ convolution block to obtain the enriched image features of the shape $c \times h \times w$. 

\subsection*{Hierarchical Dense Spatial Coordinates (HDSC) Concatenation}

To provide the model with additional spatial context of the input image crops and achieve more accurate segmentations, we propose a new module, hierarchical dense spatial coordinates (HDSC), which hierarchically concatenates dense spatial information to the image features at different decoder stages.
Inspired by CoordConv~\cite{liu2018intriguingCoordConv}, we extend this idea by incorporating depth information and hierarchically applying it across each decoder stage to add relative spatial information.
HDSC enhances the spatial precision of convolution-based decoders, allowing for improved spatial awareness and better localization due to the explicit use of spatial information. 
This results in sharper boundaries and more accurate structures, especially when segmenting complex or irregular structures, and helps recover fine-grained spatial details lost during downsampling.

\begin{figure*}[ht!] 
\begin{center} 
\includegraphics[width=0.4\linewidth]{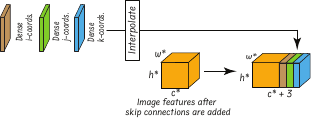} 
\caption{\label{dsp}
Concatenating dense spatial coordinates to the image features at the channel dimension after interpolation to match the dimension of the image features.
\vspace{-5mm}}
\end{center}
\end{figure*} 

For the $z^{th}$ image slice $s_z$ of dimension $H \times W$ from a volume (belonging to the age group $A_i$) with $Z$ total slices, we crop a rectangular region of size $h \times w$ with randomly selected top ($t$), bottom ($b$), left ($l$), and right ($r$) positions. 
We compute the dense coordinates as follows:

{\footnotesize{
\begin{align*}
\text{Relative $i$-Coords} = \frac{\begin{bmatrix}
l & l + 1 & \cdots & r - 1 \\
l & l + 1 & \cdots & r - 1 \\
\vdots & \vdots & \ddots & \vdots \\
l & l + 1 & \cdots & r - 1
\end{bmatrix}}{W}, \; \;
\text{Relative $j$-Coords} = \frac{\begin{bmatrix}
t & t & \cdots & t \\
t+1 & t+1 & \cdots & t+1 \\
\vdots & \vdots & \ddots & \vdots \\
b-1 & b-1 & \cdots & b-1
\end{bmatrix}}{H}, \; \;
\text{Relative $k$-Coords} = \frac{\begin{bmatrix}
z  & \cdots & z \\
\vdots & \ddots & \vdots \\
z  & \cdots & z
\end{bmatrix}}{Z}.
\end{align*}
}}

At different stages of the decoder, we interpolate the dense coordinates before concatenation to match the spatial dimensions of the features while preserving the relative spatial information. We concatenate these interpolated dense coordinates to the image features obtained after the skip connections are added (see Figure \ref{dsp}). The dense coordinates are hierarchically added at each decoder stage as shown in Figure \ref{main}.

\subsection*{Training the Conditional Segmentation Model}

We train the segmentation model on a sparse set of 2D slices which contain annotations ($<$3\% of annotated slices in a volume) using a combination of the Dice and Cross-Entropy losses. 
The Dice Loss ($\mathcal{L}_{\text{Dice}}$) is computed as:
$$\mathcal{L}_{\text{Dice}} = 1 - \frac{2 \sum_i p_i g_i}{\sum_i p_i^2 + \sum_i g_i^2},$$
where $p_i$ represents the predicted probability for the $i$-th pixel (or voxel), $g_i$ represents the ground truth label for the $i$-th pixel (with a value of 0 or 1), and the summation is performed over all pixels in the segmentation map. 
The Dice Loss effectively measures the overlap between the predicted and true masks, where a value of 0 indicates perfect overlap.
The Cross-Entropy Loss ($\mathcal{L}_{\text{CE}}$) is computed as:
$$\mathcal{L}_{\text{CE}} = - \sum_{i} g_i \log(p_i), $$
where $g_i$ is the ground truth label for the $i$-th pixel (encoded as 0 or 1), and $p_i$ is the predicted probability for the $i$-th pixel belonging to the target class. The Cross-Entropy Loss penalizes incorrect classifications, encouraging the model to produce probabilities closer to the true labels.
Thus, our overall segmentation loss $\mathcal{L}_{\text{Segmentation}}$ is taken as:
\begin{align}
\hspace{52mm} \mathcal{L}_{\text{Segmentation}} = 
\alpha \cdot \mathcal{L}_{\text{Dice}} + (1-\alpha) \cdot \mathcal{L}_{\text{CE}},
\end{align}
where the parameter $\alpha$ controls the weight of the combination.
We observed that for our dataset, equal weight for both the losses (i.e., $\alpha = 0.5$) performs the best.


\section*{Data Acquisition and Implementation Details} \label{sa_data_acq}

We use a multi-age cartilage dataset from the C57BL/6J mouse model. 
This inbred mouse strain (C57) features prominently in research modeling of typical physiological (non–pathological) conditions including skeletal health, growth, and development~\cite{zaleckas2015c57_1, lesciotto2020phosphotungstic_c57_2, lesciotto2022embryonic_c57_3}. 
The embryonic cartilaginous cranial skeleton in the C57 mouse begins to develop prior to the appearance of bone during embryogenesis and then either undergoes endochondral ossification~\cite{pitirri2023come}, disintegrates, or remains and grows in shape and size, causing morphological variations across embryonic ages.
Thus, we adopt C57 as our main data cohort due to the challenging and quick-changing cartilage and bone structures during development.

\subsection*{Dataset Details}
C57 mice were bred at both the Icahn School of Medicine at Mount Sinai and The Pennsylvania State University (PSU) via timed mating. 
Specimen production and collection took place under appropriate IACUC protocols at the specified institutions. 
Following the euthanasia of the pregnant dam, embryos were collected into cold phosphate-buffered saline (PBS) on the ice at the desired stage of embryonic development: embryonic (E) days E13.5, E14.5, E15.5, and E16.5. 
Mice litters aged less than E16 were staged using the eMOSS system10~\cite{musy2018quantitativEMOSS} to determine whether they were within $\sim$6 hours of the expected age.
All embryonic mouse specimens were stained in a 7.0\% solution of phosphotungstic acid (PTA) in 90\% methanol according to time-modified protocols~\cite{lesciotto2022embryonic_c57_3} to enhance tissue contrast during subsequent microCT scanning. 
PTA-enhanced microCT specimens were obtained at the PSU Center for Quantitative Imaging utilizing the General Electric v| tome| L300 nano/microCT system. 
The resulting raw image data were reconstructed as 32-bit and subsequently reduced to 16-bit unsigned integers before being cropped, reoriented into the coronal or transverse anatomical planes, and LUT-adjusted to obtain the 3D volumes. 

Using a previously proposed slice selection scheme \cite{zheng2019RepAnn}, representative 2D slices from C57 volumes were selected for maximum coverage within the volumes for manual cartilage annotations by experts. 
With their generous effort, we were able to obtain four partially annotated volumes for each of the four C57 ages: E13.5, E14.5, E15.5, and E16.5. 
Each volume has roughly 1600 2D slices on average with dimensions of about 1000 (height) $\times$ 1500 (width) pixels.
Within all the volumes, 40-45 2D slices (2.6\% of the total slices) were hand-annotated.  
We divided the training set into two volumes for each of the four age groups, totaling 8 volumes and 320 hand-annotated 2D slices.
Similarly, the test set consisted of two volumes per age group, a total of 8 volumes, and roughly 340 total hand-annotated 2D slices. 
We also used four volumes from mice with genetic mutations known to affect craniofacial phenotype -- denoted as Mutation A (E13.5), Mutation B (E15.5), and Mutation C (E14.5, E16.5) \cite{perrine2023embryonic} -- to study the generalizability of the conditionally trained models. 
We have in total four volumes ($\sim$1300 slices per volume) out of which 201 slices were hand-annotated for cartilage segmentations and used for testing model performance on these 3 additional mutations.

\subsection*{Implementation Details}
We demonstrate the effectiveness of our proposed components using six base segmentation models: three Transformer-based models, two Convolutional Neural Network (CNN)-based models, and one CNN-Transformer hybrid model. 
All 6 models use CNN-based decoders with skip connections from the encoder to the decoder.
Except for our modified Res2Unet model (Res2Unet*), all the other models adhere to their original implementation details as described in the respective papers. Our Res2Unet* differs from the existing implementations~\cite{res2unet1, li2022res2unet} by integrating encoder and decoder features using pixel-wise summation, rather than concatenation, at the skip connections.
Each of the six models is fine-tuned for optimal configurations (width, depth, number of blocks, etc.) on our dataset. 
ConSA maintains a hidden dimension $hid\_dim$ of 64, utilizing 4 attention heads. 
The conditional vectors, which represent age and location, also are retained as 64-dimensional vectors. To reduce the risk of overfitting, we apply random jittering to the central coordinates used for the location vector by sampling uniformly in the range [0, 0.2].
The models are implemented using PyTorch 
and MONAI
.
We train them from scratch for 700 epochs using automatic mixed precision~\cite{mixedPreTrain} on an NVIDIA A10 GPU and optimized using AdamW~\cite{loshchilov2019decoupledAdamW} with a Cosine Annealing learning rate scheduler~\cite{loshchilov2016sgdr}. 
We also apply standard regularization methods, including a 10\% dropout~\cite{srivastava2014dropout} and a weight decay of 0.001~\cite{loshchilov2019decoupledAdamW}.
The images are randomly cropped to 256$\times$256 (height $\times$ width), and a non-linear Bézier curve intensity transformation~\cite{zhou2021modelsGenesis} is applied along with random rotation, and mirroring across all the three axes. 


\section*{Experiments and Results}  

Figure \ref{MAV}-(B) shows that joint training generally improves performance compared to training the models on individual ages. 
However, due to intra-age and inter-age morphological variations, not all ages benefit equally from joint training.
In this section, we show how our proposed components help segmentation models handle these variations and improve upon the base model results. 

\subsection*{Joint Training with Conditional Components on Different DL Architectures}

\begin{table*}[ht!]
\begin{center}
\resizebox{0.75\textwidth}{!}{
\begin{tblr}{
    colspec={l | r | c c c c | l},
}
\hline
\multirow{2}{15mm}{Model} 
& \multirow{2}{*}{Parameters}
& \multicolumn{6}{c} {Trained on All Four C57 Ages}\\
\cline{3-8} 
&
&  E13.5 
&  E14.5 
&  E15.5 
&  E16.5 
& Avg.
 \\
\hline
\hline
\multicolumn{7}{l} {\textbf{Transformer Models}} 
\\
\hline
UNETR \cite{hatamizadeh2022unetr}
& 14.96 M
& 53.2	 \sd{2.0}
& 63.2	 \sd{2.2}
& 75.2	 \sd{2.5}
& 82.2 \sd{1.3}
& 68.5
\\ 
UNETR \cite{hatamizadeh2022unetr} + ConSA + HDSC
& 15.03 M
& 56.8	 \sd{1.6}
& 66.2	 \sd{2.2}
& 77.6	 \sd{2.1}
& 83.5 \sd{0.4}
& 71.0 \inc{2.5}
\\ 
\hline
ConUNETR \cite{sapkota2024conunetr}
& 14.96 M
& 55.6  \sd{1.2}	
& 65.3	 \sd{0.2}
& 77.1	 \sd{0.5}
& 83.9 \sd{0.6}
& 70.5 
\\ 
ConUNETR \cite{sapkota2024conunetr} + ConSA + HDSC
& 15.03 M
& 58.5	\sd{0.3}
& 67.6	 \sd{0.6}
& 78.4	 \sd{0.8}
& 84.9 \sd{0.5}
& 72.3  \inc{1.8}
\\ 
\hline
SwinUNETR v2 \cite{he2023swinunetrV2}
& 7.08 M
& 60.0	 \sd{1.6}
& 74.0	 \sd{1.1}
& 86.5	 \sd{0.5}
& 88.6 \sd{0.5}
& 77.3 
\\ 
SwinUNETR v2 \cite{he2023swinunetrV2} + ConSA + HDSC
& 7.16 M
& 62.3	 \sd{2.4}
& 75.5	 \sd{1.4}
& 87.2	 \sd{0.6}
& 90.1 \sd{0.8}
& 78.8 \inc{1.5}
\\ 

\hline
\hline
\multicolumn{7}{l} {\textbf{CNN or Hybrid Models}} \\
\hline
U-Net \cite{unet} 
& 7.94 M
& 61.7 \sd{2.0}	
& 76.4 \sd{0.5}	
& 84.9 \sd{0.8}	
& 90.8 \sd{0.3}
& 78.5 
\\ 
U-Net \cite{unet} + ConSA + HDSC 
& 8.04 M
& 65.6 \sd{0.7}	
& 78.1 \sd{0.9}	
& 86.8 \sd{0.6}	
& 91.5 \sd{0.2}
& 80.5 \inc{2.0}
\\
\hline
UNeXt \cite{valanarasu2022unext}
& 1.47 M
& 61.9	 \sd{0.4}
& 77.7	 \sd{0.3}
& 85.3	 \sd{0.5}
& 90.5 \sd{0.4}
& 78.9 
\\ 
UNeXt \cite{valanarasu2022unext} + ConSA + HDSC
& 1.53 M
& 63.7	 \sd{1.4}
& 77.6	 \sd{0.9}
& 84.9	 \sd{1.5}
& 91.7 \sd{0.1}
& 79.5 \inc{0.6}
\\
\hline
Res2UNet* 
& 15.60 M
& 65.4	 \sd{1.1}
& 78.7	 \sd{0.2}
& 87.6	 \sd{0.6}
& 91.8 \sd{0.1}
& 80.9
\\ 
Res2UNet* + ConSA + HDSC
& 15.88 M
& 67.2	 \sd{0.5}
& 81.2	 \sd{0.3}
& 88.2	 \sd{0.3}
& 92.3 \sd{0.1}
& 82.2 \inc{1.3}
\\  
\hline
\end{tblr}
}
\caption{\label{main_table} Segmentation performances (Dice Score) of various DL architectures, both with and without our proposed conditional components. All the models are trained on the combined training set consisting of four different ages from the C57 dataset (joint training) and evaluated on the multi-age C57 test sets. \vspace{-5mm}}
\end{center}
\end{table*}

For this study, we combine the training data from all four age groups in the C57 dataset, and evaluate the models on the corresponding test set of each age (see Table \ref{main_table}).
In the Transformer-based architectures, our proposed components yield at least a 2.3\% Dice improvement on the earliest age (E13.5) and at least a 1.0\% Dice score improvement on the most developed age (E16.5). 
On average, these components contribute to at least a 1.5\% improvement across all the ages for Transformer-based models.
Similarly, when applied to CNN-based architectures, our components deliver more than a 1.8\% improvement on the earliest age and a 0.5\% improvement on the most developed age, with an average improvement of 1.3\% across all the ages.
The hybrid model UNeXt~\cite{valanarasu2022unext}, which employs CNN-based encoders and decoders while using a tokenized multilayer perceptron (MLP) to introduce Transformer-like effects at the bottleneck, also experiences an average improvement of at least 0.6\% when enhanced with our components.
Thus, with our proposed conditional training components, all the models experience notable performance gains by effectively handling the structural variations in multi-age cartilage data while simultaneously leveraging the underlying similarities across the ages.


























\subsection*{Zero-shot Transfer}

We use three distinct mutations (Mut A, Mut B, and Mut C) that are unseen during training to evaluate how well the conditional models generalize in a zero-shot transfer scenario (i.e., without training but only testing on the new data).
Since the models were trained on cranial and post-cranial regions in the C57 dataset, we extracted only those regions from the volumes of the new mutations for testing. 
The models jointly trained across all the ages of the C57 dataset are directly applied to one volume from each of the four known ages across the three unseen mutations.
In Table~\ref{zero shot}, we show the results of the best-performing CNN- and Transformer-based models. 
The models incorporating our conditional components demonstrate superior generalizability compared to their base versions without the added components.
We achieve an average improvement of 7.6\% Dice on the Transformer-based model and 7.3\% on the CNN-based model, drastically improving the generalizability of our networks with minimal computational overhead. 

\begin{table*}[h]
\begin{center}
\resizebox{0.70\textwidth}{!}{
\begin{tblr}{
    colspec={r| l l l l |l},
}
\hline
\multirow{2}{15mm}{Model} & \multicolumn{5}{c} {Trained on All Four C57 Ages}\\
\cline{2-6}
& Mut A E13.5
& Mut B E14.5 
& Mut C E15.5
& Mut B E16.5  
& Avg.
 \\
\hline
SwinUNETR v2
& 50.4
& 71.9
& 67.1
& 72.6 
& 65.5
\\
SwinUNETR v2 + ConSA + HDSC
& 56.9 \inc{6.5}
& 78.1 \inc{6.2}
& 76.3 \inc{9.2}
& 80.9 \inc{8.3}
& 73.1 \inc{7.6}
\\
\hline
Res2UNet* 
& 48.5
& 69.5 
& 60.6
& 70.8
& 62.4
\\
Res2UNet* + ConSA + HDSC
& 53.7 \inc{5.2}
& 76.8  \inc{7.3}
& 68.2 \inc{7.6}
& 79.9 \inc{9.1}
& 69.7 \inc{7.3}
\\
\hline
\end{tblr}
}
\caption{\label{zero shot} Results of zero-shot transfer (i.e., performing inference on new data without fine-tuning) on unseen datasets. All the models are trained on the combined training set of the four ages of the C57 dataset (joint training) and applied to the test set volumes belonging to four ages of three different mutations (Mut A, Mut B, and Mut C). \vspace{-8mm}}
\end{center}
\end{table*}

\subsection*{Components and Ablation Study}

We study the optimal configurations of the proposed conditional components and their significance within the overall conditional segmentation model. For this study, we select the best-performing CNN-based model (Res2Unet*) and Transformer-based model (SwinUNETR v2). These models are jointly trained on the C57 dataset and applied to the multi-age C57 test set.
We compare the model performances trained with and without the conditional components.

\begin{table*}[h]
\begin{center}
\begin{tabular}{cc}  
\resizebox{0.345\textwidth}{!}{
\begin{tblr}{
    colspec={l|l l},
}
\hline
\multirow{2}{*}{Components} & \multicolumn{2}{c} {Trained on All Four C57 Ages}\\
\cline{2-3} 
& Res2Unet* 
& SwinUNETR v2 
\\
\hline
Base Model & 80.9 & 77.3
\\
\hline
+ FM (age) & 79.4 \dec{1.5} & 75.3 \dec{2.0}
\\
+ ConSA (age) & 81.7 \inc{0.8} & 78.2 \inc{0.9}
\\
\hline
+ FM (age+loc) & 77.8 \dec{3.1} & 75.8 \dec{1.5}
\\
+ ConSA (age+loc) & 81.8 \inc{0.9} & 78.1 \inc{0.8}
\\
\hline
+ HDSC$_{encoder}$ & 80.5 \dec{0.4} & -  \\
+ HDSC$_{encoder+decoder}$ & 80.6 \dec{0.3} & - \\
+ HDSC$_{decoder}$ & 81.6 \inc{0.7} & 77.8 \inc{0.5}
\\
\hline
\end{tblr}
}
&  
\resizebox{0.55\textwidth}{!}{
\begin{tblr}{columns={colsep=3pt},
    colspec={l l l l | c  c | c c },
}
\hline
\multirow{2}{*}{\rotatebox{45}{Decoder}} & 
\multirow{2}{*}{\rotatebox{45}{+ ConSA (Age)}} & 
\multirow{2}{*}{\rotatebox{45}{+ ConSA (Loc)}} & 
\multirow{2}{*}{\rotatebox{45}{+ HDSC}} & 
\multicolumn{2}{c}  Res2Unet* &
\multicolumn{2}{c}  SwinUNETR v2
\\
\cline{5-8} 
& & &  &  {\rotatebox{45} {Early Ages}}  & {\rotatebox{45} {Later Ages}}  &  {\rotatebox{45} {Early Ages}}  & {\rotatebox{45} {Later Ages}}  \\
\hline  
 \checkmark &&& & 72.1 & 89.7 & 67.0  & 87.5 \\
 \hline
   \checkmark & \checkmark &   &&  73.0 & 90.3 & 68.0  & 88.3 \\
   \checkmark & \checkmark   & \checkmark   &&  73.5 & 90.0 & 68.6  & 87.6 \\
    \hline
  \checkmark & \checkmark & \checkmark  & \checkmark &   74.0 & 90.1 & 68.5  & 87.9  \\
  \checkmark & \checkmark &  & \checkmark &   74.2 & 90.3 & 68.9  & 88.7  \\
\hline
\end{tblr}
}
\end{tabular}
\caption{\label{comp_ab} Left: Comparison of different components and their configurations. Right: The ablation study shows the benefits of different components when they are used. All the models are trained on the combined training set of four ages of the C57 dataset (joint training) and applied to the multi-age C57 test sets. \vspace{-5mm}}
\end{center}
\end{table*}

\subsubsection*{Feature Modulation}

In the literature, feature modulation (FM) may be considered as an alternative to conditional self-attention. Thus, we study feature modulation conditioned on age and spatial information following the widely-used Feature-wise Linear Modulation (FiLM) method~\cite{perez2018film}. 
The implementation details can be found in the Supplementary Material. 
In the two blocks below ``Base Model'' in Table~\ref{comp_ab}-Left, we compare the performances of the feature modulation and conditional self-attention layers. 
Feature modulation yields less favorable results than the base model. Despite the morphological differences in the multi-age dataset, these variations may not provide sufficiently strong learning signals for modulated features to accurately capture the structural differences. In comparison, the soft feature modulation attained through conditional self-attention is more effective in the sparse annotated data setting of our multi-age cartilage dataset.

\subsubsection*{Spatial Information in Different Stages of CNN Encoders and Decoders}
 
We investigate the optimal placement of our HDSC in the segmentation network. Note that for Transformer-based encoders, we exclude HDSC from the study, as the spatial embeddings in such an encoder architecture already capture positional relationships between tokens, making the addition of HDSC redundant. However, since state-of-the-art segmentation models typically employ CNN-based decoders, regardless of their encoder type, we focus our effort on enhancing spatial awareness in the decoder by integrating HDSC into it.
As shown in the last block of Table~\ref{comp_ab}-Left, placing HDSC in the encoder or in both the encoder and decoder leads to worse performance than the base model. However, placing HDSC in the CNN-based decoder blocks results in up to a 0.7\% improvement compared to the base model. Adding spatial information in the encoder can hinder learning in sparse annotated data settings by possibly causing the model to overfit to specific spatial cues, rather than focusing on general and robust features. In contrast, introducing spatial information in the decoder helps the model better localize and refine object boundaries, and improves segmentation precision by enhancing the high-level features and making them more effective in scarce annotated data scenarios.


\subsubsection*{Ablation Study}

Table~\ref{comp_ab}-Right presents the segmentation performances with various components in combination. The results are obtained by training the models on all four C57 ages and evaluating on the corresponding C57 test set. Average Dice scores for the early ages (E13.5 + E14.5) and later ages (E15.5 + E16.5) are reported for both the best-performing CNN- (Res2UNet*) and Transformer-based (SwinUNETR v2) models. Incrementally adding the proposed components improves performance for both the CNN and Transformer models across both the early and later age groups. Notably, for the later ages, location information during ConSA is less critical, as HDSC alone can capture the necessary spatial information due to the more developed cartilage structures in these ages. The best results are achieved by combining ConSA and HDSC, leading to performance gains of up to 2.1\% for the early ages and up to 1.2\% for the later ages while adding merely at most 0.3 million parameters.


\section*{Discussion}

%

%



\begin{figure*}[t] 
\begin{center} 
\includegraphics[width=0.9\textwidth]{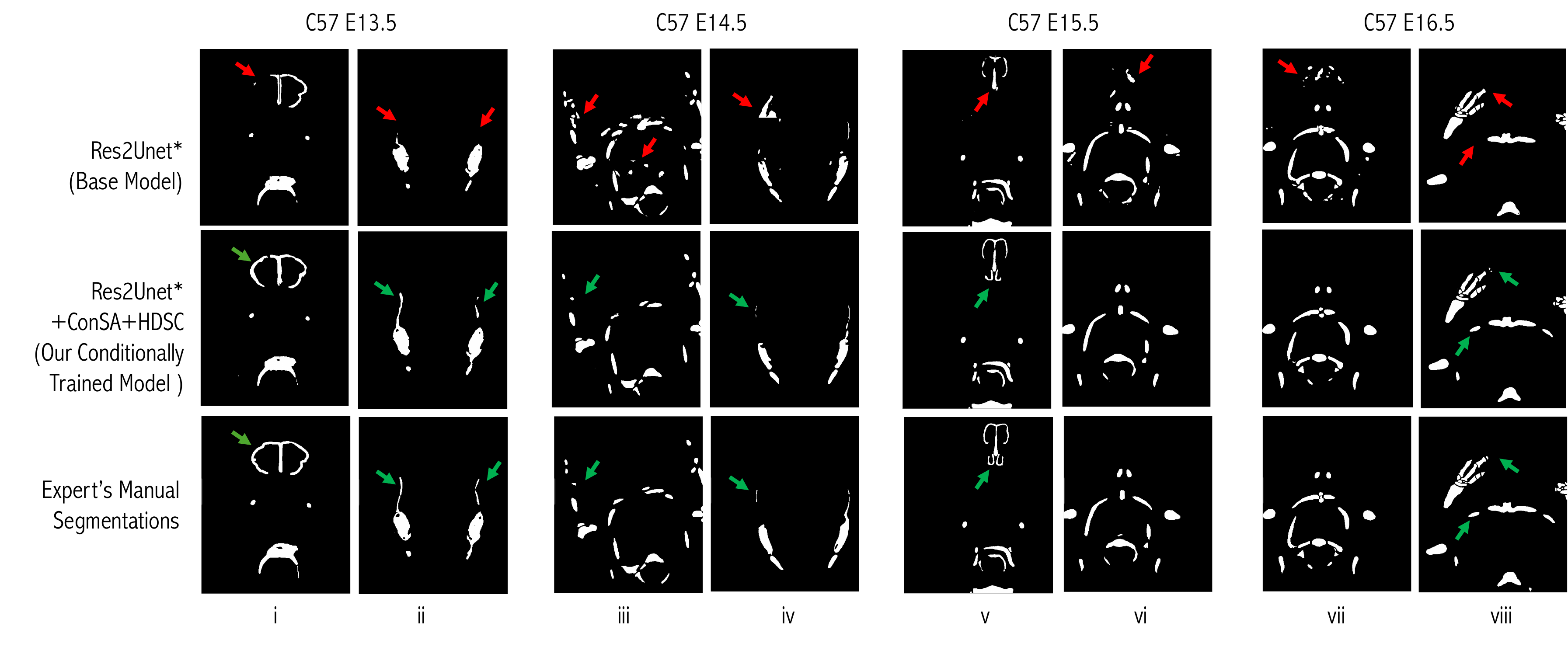} 
\caption{\label{quals} Visual examples of segmentation results (joint trained on the C57 multi-age training set and applied to the C57 multi-age test set) obtained by the Res2Unet* model (top row), Res2Unet* + ConSA + HDSC (middle row), and expert's manual segmentations = ground truth (bottom row). Red arrows indicate poorly segmented or ``noisy'' output, while green arrows point to expected or well-segmented regions. \vspace{-8mm}
}
\end{center}
\end{figure*} 

\begin{figure*}[t] 
\begin{center} 
\includegraphics[width=0.6\linewidth]{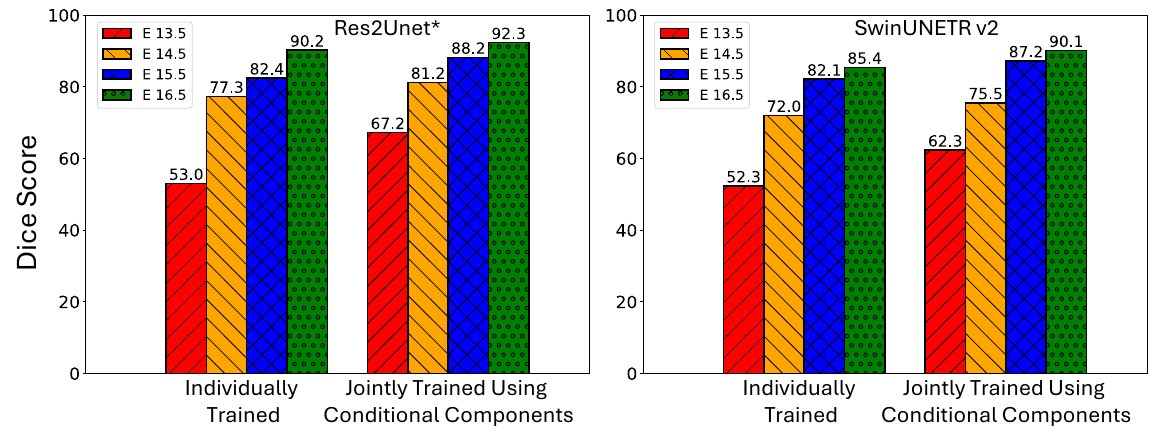} 
\caption{\label{increase_bargraph}
Comparison of segmentation results (Dice scores) using individually trained models versus joint training with our proposed conditional components. Left: The Res2Unet* model. Right: The SwinUNETR v2 model. The models are trained using the C57 multi-age training set and applied to the C57 multi-age test set. \vspace{-8mm}
}
\end{center}
\end{figure*}

From an anatomical perspective, our conditional components considerably enhance the segmentation accuracy in several 
aspects. Figure~\ref{quals} presents some visual segmentation examples of the Res2UNet* models jointly trained on all four age groups of the C57 dataset and applied to the C57 multi-age test set. The conditionally trained Res2UNet* model (Res2Unet* + ConSA + HDSC) shows noticeable improvements compared to its base version, generating segmentations with fewer background artifacts, better structural consistency, and improved recognition of the anatomical details of the target cartilage.
For the earliest age group (E13.5), the base model (Res2Unet*) fails to fully capture the circular shape of the nasal capsule (the first row of Figure~\ref{quals}-i) and is unable to segment the thin superior portion of the lateral walls of the chondrocranium (the first row of Figure~\ref{quals}-ii). In contrast, the conditionally trained model better captures the true morphology of these tissues, preserving not only the shape but also the continuity of the structure (the second row of Figure~\ref{quals}-i and -ii) and closely mimicking the ground truth (the third row of Figure~\ref{quals}-i and -ii) for the cartilage segmentation produced by the anatomical expert (hand annotation).
Similarly, for the age groups E14.5 and E15.5, the segmentations by the conditionally trained model are significantly cleaner, with less ``noise'' (fewer scan artifacts and scattered voxels). The boundaries around the structures are better identified, and the unwanted artifacts present in the base model's segmentations are substantially reduced, resulting in smoother segmented surfaces. For E14.5, costal and vertebral cartilages are segmented more accurately, reducing ``noise'' output (the second row of Figure~\ref{quals}-iii) and eliminating over-segmentation (false positives) on the lateral walls of the chondrocranium compared to the base model (the second row of Figure~\ref{quals}-iv).
Moreover, for the latest age group (E16.5), the segmentation outputs of the base model appear fragmented and incomplete (the first row of Figure~\ref{quals}-viii), with some areas showing false positives (the first row of Figure~\ref{quals}-vii). In contrast, our model captures these regions with enhanced accuracy, providing a more consistent, detailed, and anatomically correct segmentation even for small skeletal elements, such as individual phalanges within the extremities of the forelimbs. Likewise, our model fully segments the sternum (the second row of Figure~\ref{quals}-viii) while the base model presents fragmented elements that lack sharpness and contain several false negatives (the first row of Figure~\ref{quals}-viii). The conditionally trained model distinguishes individual skeletal elements in the extremities, correctly identifying and preserving their true morphology as compared to the expert's manual annotations.

Figure~\ref{increase_bargraph} further reports quantitative improvements of our model over the ones trained on single age groups separately.
In Figure~\ref{increase_bargraph}, ``Individually Trained" represents results when the models are trained individually on a specific age (e.g., E13.5) and are applied to the same age (e.g., E13.5); ``Jointly Trained using Conditional Components" refers to the models trained jointly on all the C57 ages and applied to all the C57 ages (E13.5, E14.5, E15.5, and E16.5). 
Compared to the common practices of individually training the models on each age group, our proposed idea of joint training on all ages using the conditional components with CNN-based (Transformer-based) models yields a 14.2\% (10.0\%) improvement on the E13.5 age during which the cartilage structures are far from fully developed, and 2.1\% (4.7\%) on the E16.5 age when the cartilage structures are mostly developed. 


\section*{Conclusions}  
\label{sec:5}

In this paper, we addressed the challenges of structural variations in multi-age embryonic cartilage segmentation in 3D micro-CT images by exploring conditional modules for DL segmentation models with our new UniCoN approach.
Our proposed conditional components enabled joint training on multi-age data, reducing annotation and training efforts. These proposed encoder-agnostic conditional components are universally compatible with CNN-based, Transformer-based, and hybrid CNN-Transformer-based segmentation models with merely a very small increase in model complexity. Several conditional configurations, integrating age-specific and spatial information into the decoding process, were evaluated, and the experimental results showed that the models trained using our components (e.g., Res2Unet* + ConSA + HDSC) yield considerable Dice score improvements across all ages compared to the previous state-of-the-art conditional model, ConUNETR. On unseen data, the models trained using our components resulted in 7.5\% Dice improvement. These findings pave the way for further research into robust, universal segmentation models capable of handling diverse datasets even when limited annotations are available.

\section*{Data Availability} 
The datasets used in this manuscript are partially available for public use, and the dataset details can be found in our previous publication ~\cite{perrine2023embryonic} through this 
\href{https://www.datacommons.psu.edu/commonswizard/MetadataDisplay.aspx?Dataset=6367}{link}.
\vspace{-2mm}
\bibliography{sample}
\newpage
\section*{Supplementary Material} \label{supp}

\subsubsection*{Details on Feature Modulation}

As an alternative conditioning scheme, we apply Feature Modulation (FM) conditioned on age and spatial information, following the widely-used Feature-wise Linear Modulation (FiLM) method~\cite{perez2018film} without the RNN component. We first use linear layers to map the age and spatial information to obtain age and location vectors
(refer to Figure~\ref{conSA}-(A)). 
Next, we concatenate these vectors into one vector and map it through an MLP to obtain the feature modulation parameters $\gamma$ and $\beta$, as: 
\begin{align}
\gamma_{i,c}, \; \beta_{i,c} &= MLP_F \; (concat\; [Age \;Vector, \; \; Location \; Vector]).
\end{align}
We then perform a linear transformation of each image feature ($F_{i,c}$) using the modulation parameters, as: 
\begin{align}
FM(F_{i,c} | _{\gamma_{i,c}, \beta_{i,c}} = \gamma_{i,c} \cdot F_{i,c} + \beta_{i,c}).
\end{align}
We place the FM blocks at every decoder stage after each block's last convolution (and Batch Normalization if used). 
\begin{figure*}[h!] 
\begin{center} 
\includegraphics[width=0.7\textwidth]{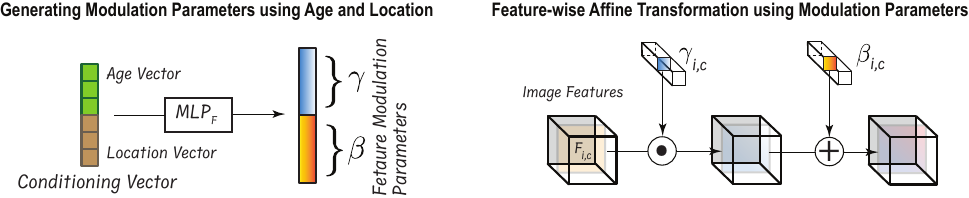} 
\caption*{\label{moreTraningData} \textbf{Supplementary Figure 1}: 
Illustrating feature modulation conditioned on age and spatial information. 
}
\end{center}
\end{figure*}

\end{document}